\documentclass{emulateapj}
\usepackage{graphicx} \usepackage{color} \usepackage{natbib}

\usepackage{amsmath,amssymb,amsfonts}
\usepackage{epsf}
\usepackage{epsfig}
\usepackage{longtable}

\newcommand{\lrr}{Living Rev. Rel.}

\shorttitle{Nucleosynthesis constraints on the neutron star-black hole merger rate}
\shortauthors{Bauswein, Ardevol Pulpillo, Janka, and Goriely}

\begin{document}

\title{Nucleosynthesis constraints on the neutron star-black hole merger rate}

\author{A.~Bauswein\altaffilmark{1}, R.~Ardevol Pulpillo\altaffilmark{2,3}, H.-T.~Janka\altaffilmark{2}, and  S.~Goriely\altaffilmark{4}}

\altaffiltext{1}{Department of Physics, Aristotle University of Thessaloniki, GR-54124 Thessaloniki, Greece}
\altaffiltext{2}{Max Planck Institute for Astrophysics, Karl-Schwarzschild-Str.~1, 85748 Garching, Germany}
\altaffiltext{3}{Physik Department, Technische Universit\"at M\"unchen, James-Franck-Str.~1, 85748 Garching, Germany}
\altaffiltext{4}{Institut d'Astronomie et d'Astrophysique, Universit\'e Libre de Bruxelles, C.P. 226, B-1050 Brussels, Belgium}

\begin{abstract}
We derive constraints on the time-averaged 
event rate of neutron star-black hole (NS-BH) mergers
by using estimates of the population-integrated production of heavy 
rapid neutron-capture (r-process) elements with nuclear mass numbers
$A > 140$ by such events in comparison to 
the Galactic repository of these chemical species. 
Our estimates are based on relativistic hydrodynamical
simulations convolved with theoretical predictions of 
the binary population. This allows us to determine a strict upper limit of
the average NS-BH merger rate of $\sim$6$\times 10^{-5}$
per year. We quantify the uncertainties of this estimate to be within
factors of a few mostly because of the unknown BH spin distribution of such
systems, the uncertain equation of state of NS matter, and
possible errors in the Galactic content of r-process material. Our approach
implies a correlation between the merger rates of NS-BH
binaries and of double NS systems.
Predictions of the detection rate of gravitational-wave signals from
such compact-object binaries by Advanced LIGO and
Advanced Virgo on the optimistic side are incompatible with the
constraints set by our analysis.
\end{abstract}


\section{Introduction}
To date no neutron star-black hole (NS-BH) binaries are known. 
Such compact object binaries, however, emit
gravitational-waves (GWs), which for some systems leads to the merging of
the binary components within the Hubble time. The inspiral and coalescence
of NS-BH and NS-NS binaries are among the primary targets for the upcoming
Advanced LIGO and Advanced Virgo GW detectors \citep{2010CQGra..27q3001A}.
Compact object mergers (COMs) that lead to a post-merging accretion torus around
the relic BH are also promising progenitor candidates of short-hard gamma-ray
bursts (GRBs)~\citep{1986ApJ...308L..43P,1989Natur.340..126E,2013arXiv1311.2603B}.
Some fraction of the stellar matter is expected to become gravitationally
unbound during and after the binary collision and is likely to undergo the rapid
neutron-capture process
(r-process) \citep{1974ApJ...192L.145L,1977ApJ...213..225L}, 
creating neutron-rich elements whose astrophysical production site
has not been unambiguously identified yet. Evidence from chemogalactic evolution
studies grows that COMs could be the dominant sources of the Galactic heavy
r-process nuclei 
\citep[traced by the enrichment history of europium;][and references therein]{2013arXiv1307.0959M,2014arXiv1407.3796S,2014arXiv1407.7039V}. 
The radioactive decays of
r-process material may power the thermal emission of a detectable
electromagnetic counterpart of the
merger~\citep{1998ApJ...507L..59L,2010MNRAS.406.2650M}.
Tentative evidence for such an event has been reported
by~\citet{2013ApJ...774L..23B,2013Natur.500..547T}.

Because of the lack of observational data, predictions of the
frequency of NS-BH mergers rely on theoretical studies
\citep[e.g.][]{1993ARep...37..411T,2003MNRAS.342.1169V,2008ApJ...672..479O,2012ApJ...759...52D,2013arXiv1307.0959M,2013ApJ...779...72D,2014arXiv1403.4754P}. These
population synthesis models estimate Galactic merger rates between
$2\times 10^{-9}$ and $10^{-5}$ per
year~\citep{2014arXiv1403.4754P}. The range reflects the challenge in
comprehensively modelling the formation and evolution of stellar
binaries and their remnants.

In this work we determine an independent upper limit on the merger rate
of NS-BH binaries by comparing the predicted r-process nucleosynthesis
yields of such systems with the observed Galactic amount of r-process
material. Similar arguments were used,
e.g., in \citet{1974ApJ...192L.145L,1999ApJ...525L.121F,2000ApJ...534L..67Q,2010MNRAS.406.2650M,2011ApJ...738L..32G,2012MNRAS.426.1940K,2013MNRAS.430.2585R,2013ApJ...773...78B,2014arXiv1401.2166P}
mostly in the context of NS-NS mergers or COMs in
general. Here we elaborate on these estimates by employing population
integrated yields, which is crucial for NS-BH systems because of the
strong dependence on binary parameters. We use ejecta masses
determined by relativistic simulations of relevant NS-BH systems
and, in particular, quantify the uncertainties of our estimates.
Information on the
hydrodynamical and nucleosynthesis calculations of NS-BH systems is
provided in Sect.~\ref{sec:model}. Our rate estimates are discussed in
Sect.~\ref{sec:rate} and combined with NS-NS mergers in
Sect.~\ref{sec:nsns}. We finish with conclusions in Sect.~\ref{sec:sum}.

\section{Numerical modeling}\label{sec:model}
We simulate NS-BH mergers with a relativistic smooth particle
hydrodynamics (SPH) code, which evolves the hydrodynamical quantities
comoving with the fluid \citep{2002PhRvD..65j3005O,BausweinThesis,2014arXiv1406.2687J}.
The Einstein equations are solved by imposing the conformal flatness
condition (CFC) on the spatial metric as formulated in \citet{2009PhRvD..79b4017C}. 
This allows us to
model the spinning and moving BH by the ``static puncture''
approach~\citep{1997PhRvL..78.3606B}, which is commonly used to
construct initial data for fully relativistic
calculations~\citep{2011LRR....14....6S}. (Despite the term ``static
puncture'' the BH carries linear and angular momentum and moves during
the simulations~\citep{BausweinThesis,2011LRR....14....6S}.) To this end we 
replace the metric equations in our hydrodynamics
solver of~\citet{2002PhRvD..65j3005O} by Eqs.~(59)--(62)
of \citet{2011LRR....14....6S}, using their definitions of
Eqs.~(56)--(58) for treating BH spin and momentum. Our existing multigrid
solver then yields the metric for NS-BH systems within the CFC approximation.

\begin{figure}
\includegraphics[height=8.6cm,angle=270]{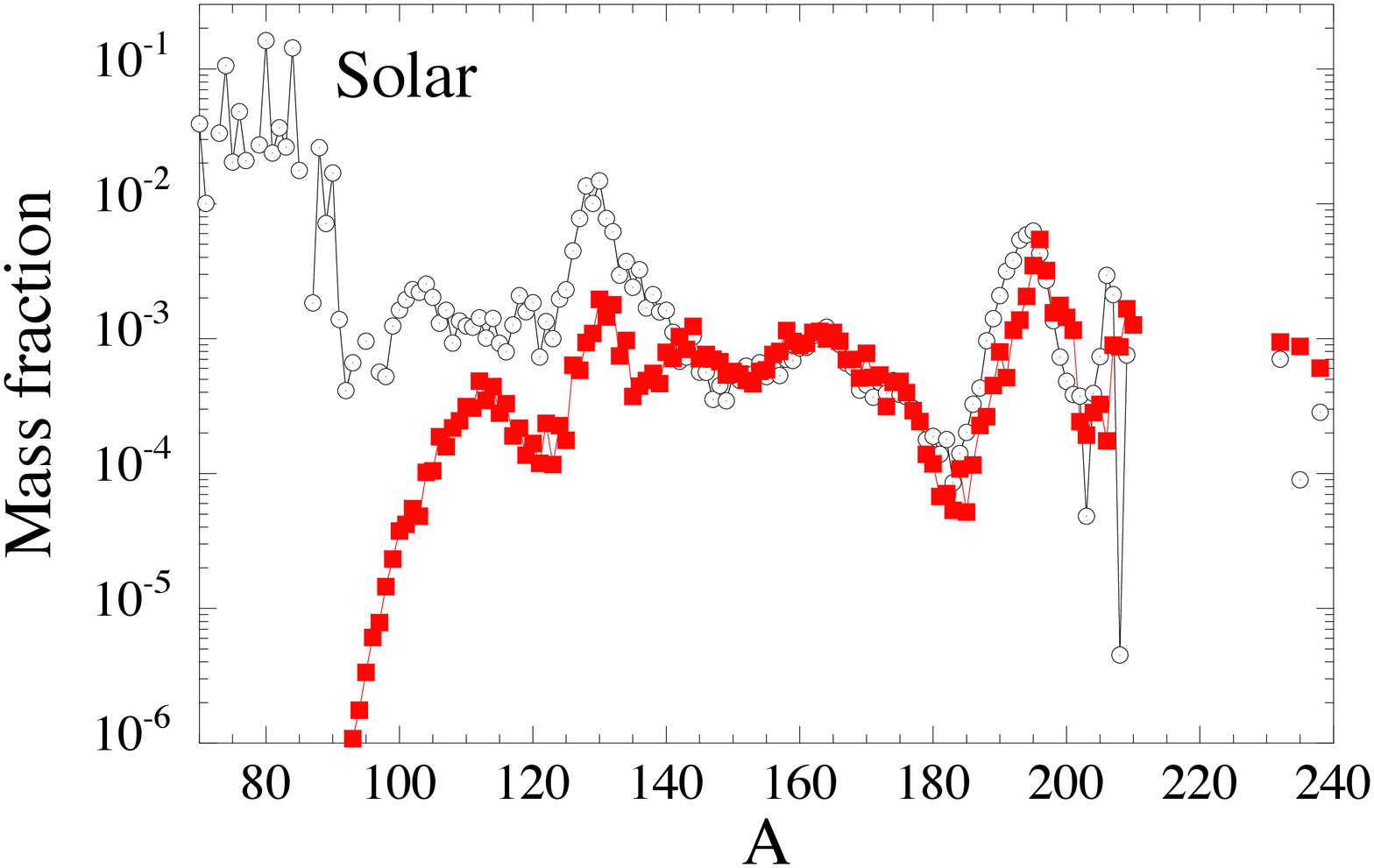}
\caption{\label{fig:ab} Abundance distribution vs.\ atomic
  mass for merger ejecta of a 1.35\,$M_\odot$ NS and a
  7\,$M_\odot$ BH with $a_\mathrm{BH}=0.5$ (red symbols) 
 normalised to the solar distribution (black symbols; normalized
  to unity) in the $A=165$ rare-earth region.}
\end{figure}

The simulations start from circular quasi-equilibrium orbits a few
revolutions before merging. The NSs are initially
nonrotating, while we
investigate different initial BH spins perpendicular to the orbital
plane. The NS matter is initially at zero temperature and in
neutrinoless beta-equilibrium. We employ the microphysical DD2
equation of state
(EoS)~\citep{2010NuPhA.837..210H}, which is a
moderately stiff EoS and yields a NS radius of 13.21\,km for a
1.35\,$M_{\odot}$ NS.  The fluid is modelled with a resolution of
$\sim$170,000 SPH particles. We test different resolutions to confirm that
our results are converged, in particular the ejecta mass is determined
with an accuracy of $\sim$30\%.

Throughout this paper we quantify the (initial) BH spin by the dimensionless
spin parameter $a_{\mathrm{BH}}=J_{\mathrm{BH}}/M_{\mathrm{BH}}^2$
with the BH angular momentum $J_{\mathrm{BH}}$ and gravitational
mass $M_{\mathrm{BH}}$. Masses refer to the gravitational mass of the
object (for binaries at infinite separation).

For fixed NS mass of $M_{\mathrm{NS}}=1.35~M_{\odot}$ we compute the ejecta masses for systems with initial BH masses of 5, 7, and
10\,$M_{\odot}$ and initial BH spins varied between 0.0 and 0.9 (see
Table~\ref{tab:ejecta}). The initial spin $a_{\mathrm{BH}}$
has the strongest impact on
the ejecta mass. For $M_{\mathrm{BH}}=5\,M_{\odot}$ 
$M_{\mathrm{ej}}$ grows by a factor of $\sim$200 when $a_{\mathrm{BH}}$
is increased from zero to 0.9. For systems with the
same $a_{\mathrm{BH}}$, the ejecta mass shows only a moderate variation 
with increasing $M_{\mathrm{BH}}$ until $M_{\mathrm{ej}}$ 
starts to plummet at higher BH masses.
This behavior is understandable because with increasing $M_{\mathrm{BH}}$
the radius of the innermost stable orbit grows faster than the tidal
disruption radius until the tidal disruption of the NS changes to a
plunge at higher $M_{\mathrm{BH}}$.
The steep reduction of $M_{\mathrm{ej}}$ occurs at higher BH mass for higher
initial spin.  Our results are qualitatively and
quantitatively consistent with the findings
of~\citet{2013ApJ...778L..16H,2013PhRvD..87h4006F,2013PhRvD..88d1503K,2014arXiv1405.1121F}.
For instance in~\citet{2013PhRvD..88d1503K}, a sequence with fixed 
$a_{\mathrm{BH}}$ of 0.75 yields ejecta masses of 0.04--0.05\,$M_{\odot}$
for BH masses between 4.05 and 9.45\,$M_{\odot}$ with the H4 EoS. The
ejecta masses agree very well with our results for the DD2 EoS,
which yields NS radii comparable to those of the H4 EoS. The
simulations of~\citet{2014arXiv1405.1121F} with a somewhat softer EoS (LS220) find ejecta masses between $\sim$0.04\,$M_{\odot}$ and 
$\sim$0.15\,$M_\odot$ for systems involving BHs with spins
$a_{\mathrm{BH}} = 0.7$--0.9 and NSs with $M_{\mathrm{NS}}=1.2$--1.4~$M_{\odot}$, again in very good agreement with our computations.

Following~\citet{2011ApJ...738L..32G} we perform nucleosynthesis
calculations for a selected subset of models (detailed paper in
preparation). About 75\% of the ejecta produce heavy r-process
elements with mass numbers $A\gtrsim 140$ in close similarity
to the solar r-process abundance distribution (Fig.~\ref{fig:ab}).

\begin{table}
 \begin{ruledtabular}
\caption{\label{tab:ejecta}Ejecta masses}
\begin{tabular}{l | c c  c}
\tableline \tableline $a_{\mathrm{BH}}$ \textbackslash
$M_{\mathrm{BH}}$ & 5~$M_{\odot}$ & 7~$M_{\odot}$ & 10~$M_{\odot}$
\\ \hline 0 & 0.0004~$M_{\odot}$ & $\lesssim 2\times
10^{-6}$~$M_{\odot}$ & $\lesssim 2\times 10^{-6}$~$M_{\odot}$ \\ 0.5 &
0.042~$M_{\odot}$ & 0.0090~$M_{\odot}$ & 0.0018~$M_{\odot}$ \\ 0.7 &
0.067~$M_{\odot}$ & 0.070~$M_{\odot}$ & 0.073~$M_{\odot}$ \\ 0.9 &
0.096~$M_{\odot}$ & 0.087~$M_{\odot}$ & 0.086~$M_{\odot}$
\\ \tableline
 \end{tabular}
\tablecomments{NS-BH mergers with initial BH mass $M_{\mathrm{BH}}$,
  initial BH spin $a_{\mathrm{BH}}$, NS mass 1.35~$M_{\odot}$, and DD2 EoS.}
 \end{ruledtabular} 
\end{table}

\section{Rate estimates}\label{sec:rate}

The Galactic r-process material is estimated from the
total Galactic baryon mass, $M_{\mathrm{Gal}}\simeq 6 \times
10^{10}~M_{\odot}$~\citep{2011MNRAS.414.2446M}, and
the solar system r-process mass fraction, assuming the latter
is representative for the whole Galaxy (as suggested by the small
scatter of the europium abundance in the present-day Milky Way). Here, 
we only consider r-process elements with $A >
140$, which can be produced significantly by NS-BH mergers and in
proportions close to the solar abundances (cf.\
Fig.~\ref{fig:ab}). Including elements
with $A<140$ would require to add the contribution from
the NS-BH merger remnants and would increase the uncertainties in the
total ejecta mass \citep[see][]{2014arXiv1406.2687J}.

Decomposing the solar-system 
abundances of elements heavier than iron 
into their s- and r-process components
\citep{1999AA...342..881G}, the 
total mass fraction of r-process elements with $A > 140$
in the solar system is estimated as $X^{\odot}_{r,A>140}=3.1\times
10^{-8}$. This implies a Galactic content of such material of
$\sim$$M_{r,A>140}=M_{\mathrm{Gal}} \times
X^{\odot}_{r,A>140} =1860~M_{\odot}$.

The r-process inventory of $M_{r,A>140}$ has to be compared with the
yield of $A>140$ material (75\% of the total
ejecta) from the NS-BH population over the Galactic history of
$T_{\mathrm{Gal}}=10^{10}$~yrs. We estimate the
production by NS-BH mergers by convolving the ejecta mass in
dependence on the binary parameters with the binary
distribution provided by population synthesis or deduced from
observations.

Depending on the metallicity and other model assumptions, 
population synthesis studies
by \citet{2012ApJ...759...52D}\footnote{www.syntheticuniverse.org}
predict average BH masses of 8--10\,$M_{\odot}$, roughly consistent with
observations \citep{2013SSRv..tmp...73M}. For
simplicity and independence of additional modeling assumptions
(like delay-time distribution and metallicity evolution), we
consider only the time-averaged merger rate over the Galactic history.
We exclude binaries with lifetimes exceeding the age of the Galaxy
and checked that our estimates do not depend significantly on the
lifetimes of the merging systems. Our constraints yield
firm upper limits of the present-day NS-BH merger rate.

The merging binaries contain NSs with an average mass of about
1.5\,$M_\odot$, while for computational reasons and for a better
comparison with the literature we employ a NS mass of
1.35\,$M_{\odot}$ compatible with earlier population synthesis
models \citep{2008ApJ...682..474B} and NS masses observed in double
NS binaries \citep{2012ApJ...757...55O}. This assumption may
introduce an uncertainty of a factor of two \citep{2014arXiv1405.1121F}.
The distribution of BH spins is not available from the adopted
population synthesis models. Hence, we treat the average BH spin as a
free parameter. The BH spin may be misaligned
with the rotation axis of the binary. About 30--80\% of the
binaries might have misalignments $>$30$^\circ$
\citep{2000ApJ...541..319K}. \citet{2008ApJ...680.1326R} showed
that for tilts below 30$^\circ$ the ejecta mass is practically
identical to the aligned case, whereas the ejecta vanish for tilt
angles exceeding 45$^\circ$. Here we adopt a fraction 
$f_{\mathrm{tilt}}=0.5$ of binaries with tilts sufficiently small to
eject matter as the aligned case. The remaining fraction of systems
is assumed to yield no ejecta. Our value of
$f_{\mathrm{tilt}}$ is compatible with findings
of \citet{2008ApJ...682..474B}, and $f_{\mathrm{tilt}}$ is expected
to vary at most between 0.2 and 0.8.

\begin{figure}
\includegraphics[width=8.6cm]{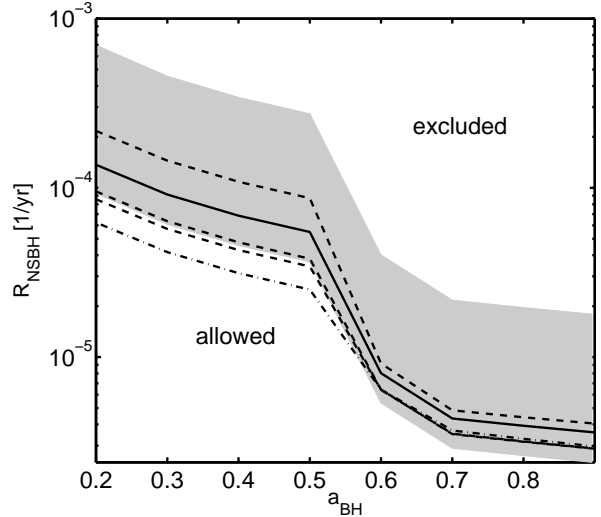}
\caption{\label{fig:rate} Upper limit on the Galactic NS-BH merger
  rate as function of BH spin for different population synthesis 
  models of \citet{2012ApJ...759...52D}. The solid line corresponds
  to their standard model (``submodel A'' at 1/10 solar metallicity).
  Dashed lines show variations to the standard model.
  The shaded region indicates uncertainties because of the unknown NS EoS.
  A rate estimate entirely based on the observed BH mass distribution
  is given by the dash-dotted line.}
\end{figure}

Using these assumptions the upper limit on the NS-BH merger rate is estimated by
\begin{equation}
R_{\mathrm{NSBH}}=\frac{M_{r,A>140}}{0.75\widetilde{M}_{\mathrm{ej}}(a_{\mathrm{BH}})
  T_{\mathrm{Gal}} f_{\mathrm{tilt}}}.
\end{equation}
$\widetilde{M}_{\mathrm{ej}}$ denotes the population averaged
ejecta mass per merger event multiplied by 0.75 to consider
only ejecta with $A>140$. Figure~\ref{fig:rate} shows
the rate as function of $a_{\mathrm{BH}}$ (solid line) for the
standard model of \citet{2012ApJ...759...52D} (their ``submodel'' A)
at 1/10 solar metallicity, where most of the mergers are
expected to take place \citep{2013ApJ...779...72D}. We find a strong
BH spin dependence. For slowly spinning BHs with
$a_{\mathrm{BH}}=0.1$ the merger rate can be as high as $2.7\times
10^{-4}\,\mathrm{yr}^{-1}$, whereas rapid rotation with 
$a_{\mathrm{BH}}=0.9$ results in a much lower rate of
$3.6\times 10^{-6}\,\mathrm{yr}^{-1}$.

The BH spin is likely to be considerably higher than 0.1 because it is
determined by the initial spin after the supernova and the
subsequent accretion in the binary
system. In \citet{2005ApJ...632.1035O} mass transfer in binaries was
found to be strong and thus to lead to generally high spins between 0.4
and 0.9 and spin-orbit alignment. In \citet{2008ApJ...682..474B} the
accretion during the binary evolution increases the initial BH spin
only moderately implying that the initial spin is the most
important parameter. Observations of X-ray binaries suggest that most
BHs have spin parameters exceeding 0.5 with an average $a_{\mathrm{BH}}$
of $\sim$0.6 \citep{2013SSRv..tmp...73M}. These spins correspond to
upper limits of the NS-BH merger rate of 
$\sim$$6\times 10^{-5}\,\mathrm{yr^{-1}}$ and
$8\times 10^{-6}\,\mathrm{yr^{-1}}$, respectively. 
The latter value is roughly compatible with the rate of
of $3.4\times 10^{-6}\,\mathrm{yr^{-1}}$ predicted by the standard
model of \citet{2012ApJ...759...52D}.

For $M_{\mathrm{BH}}/M_{\mathrm{NS}}=3$ \citet{2013PhRvD..88d1503K} and \citet{2013ApJ...778L..16H}
showed that for soft (stiff) EoSs the ejecta masses can be a factor 5 
lower (1.5 higher) compared to those for the H4 EoS, which is similar
to our DD2 EoS. The corresponding range of uncertainty is indicated
by grey shading in Fig.~\ref{fig:rate}. 
This variation is consistent with the EoS
effects quantified for ideal fluid EoSs by \citet{2013PhRvD..87h4006F}
for another binary setup. 
The relatively high ejecta masses found
by \citet{2014arXiv1405.1121F} for the soft LS220 EoS suggest that
a factor of 5 might overestimate changes associated with very 
soft EoSs. Moreover, theoretical arguments favor a nuclear EoS 
comparable to or slightly softer than DD2~\citep{2013ApJ...773...11H}.

Variants of the standard model of \citet{2012ApJ...759...52D} for
different metallicities (e.g., dashed lines in Fig.~\ref{fig:rate} 
for solar metallicity and/or ``submodel'' B) yield NS-BH merger rates 
between $2.4\times 10^{-5}\,\mathrm{yr}^{-1}$ and
$8.7\times 10^{-5}\,\mathrm{yr}^{-1}$ for $a_\mathrm{BH} = 0.5$.
The dash-dotted line in 
Fig.~\ref{fig:rate} provides a rate limit independent of
population synthesis studies by using the observed BH mass
distribution of \citet{2012ApJ...757...55O}
with a median at 7.5\,$M_{\odot}$. Because of the lower mean BH mass,
this case yields slightly lower rates than the population models.
Again, for an average $a_{\mathrm{BH}}>$0.5 this rate limit is roughly
compatible with the merger rate derived from population synthesis
at 1/10 solar metallicity \citep{2012ApJ...759...52D}.

\section{Constraints from NS-NS binaries}\label{sec:nsns}

Most population models predict NS-NS mergers to be more frequent than
NS-BH collisions~\citep{2014arXiv1403.4754P}.  The average ejecta mass
of NS-NS mergers depends strongly on the high-density EoS and is
typically smaller than that of NS-BH mergers. In contrast to NS-BH
mergers, double NS collisions produce more ejecta for soft
EoSs~\citep{2013ApJ...773...78B,2013ApJ...778L..16H}.  
In this case the population
averaged ejecta mass may be $\widetilde{M}_{\mathrm{ej,NSNS}}\approx
10^{-2}\,M_{\odot}$, whereas stiff EoSs lead to
$\widetilde{M}_{\mathrm{ej,NSNS}}\approx
10^{-3}\,M_{\odot}$; for the DD2 EoS
$\widetilde{M}_{\mathrm{ej,NSNS}}$ amounts to $\sim$$3\times
10^{-3}~M_{\odot}$~\citep{2013ApJ...773...78B}.

\begin{figure}
\includegraphics[height=8.6cm,angle=270]{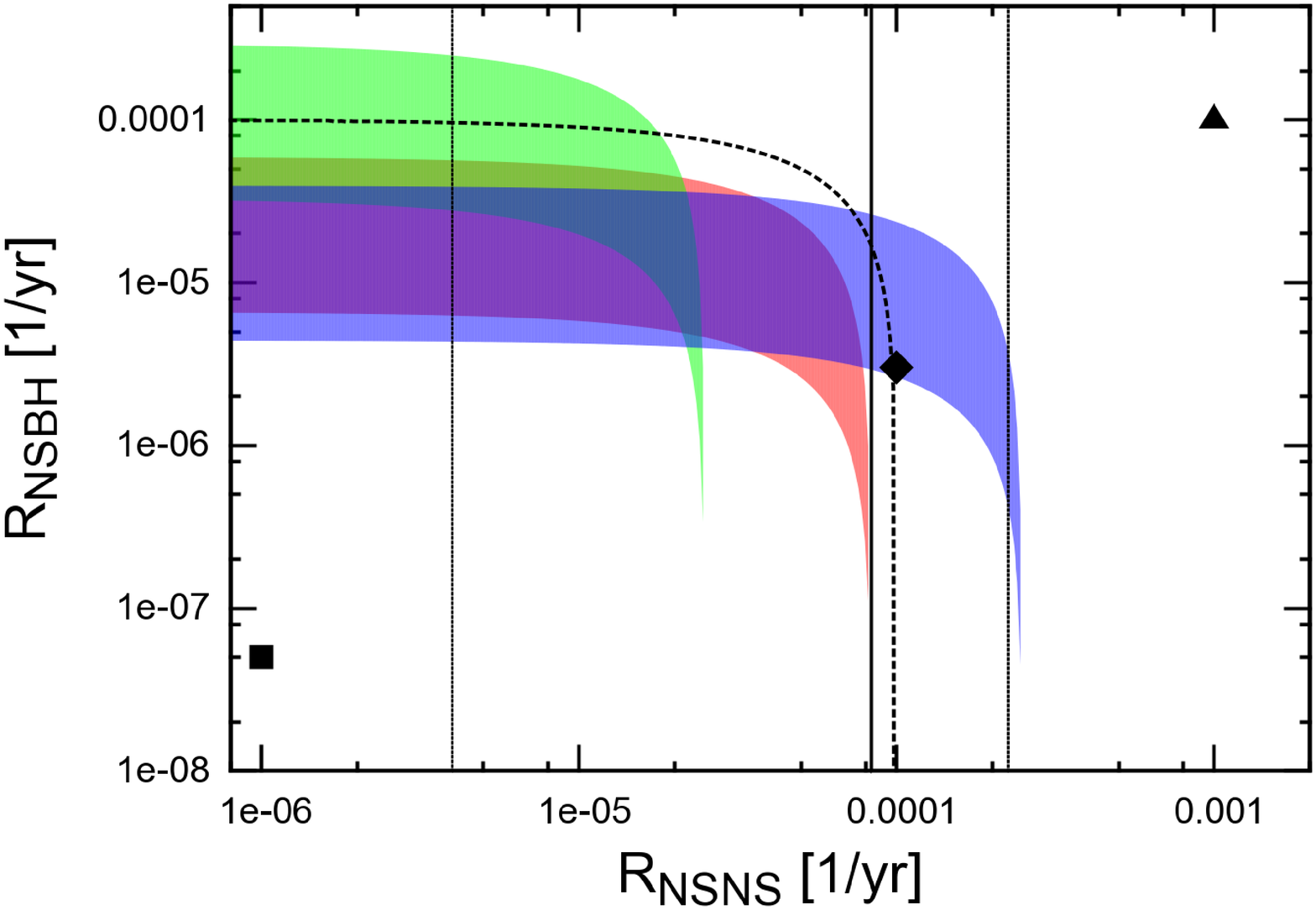}
\caption{\label{fig:cor}Average Galactic NS-BH merger rate
  vs.\ NS-NS merger rate for the DD2 EoS (reddish
  band), a stiff EoS (bluish band), and a soft EoS (greenish
  band). The widths of the bands represent variations with average BH spins
  between 0.5 and 0.7. Symbols indicate ``optimistic'' (triangle),
  ``realistic'' (diamond) and ``pessimistic'' (square) merger rates as
  compiled in \citet{2010CQGra..27q3001A}. Vertical lines mark NS-NS
  merger rates inferred from binary NS
  observations with uncertainty range \citep{2004ApJ...614L.137K}. The dashed line
  approximates a rate estimate derived from observed GRBs
  \citep{2013arXiv1311.2603B}.}
\end{figure}

Quantifying the contribution from NS-NS mergers by a time-averaged
merger rate $R_{\mathrm{NSNS}}$, the upper limit on the average Galactic 
rate for NS-BH mergers reads
\begin{equation}\label{eq:cor}
R_{\mathrm{NSBH}}=\frac{M_{r,A>140}-0.75
  \widetilde{M}_{\mathrm{ej,NSNS}} T_{\mathrm{Gal}} R_{\mathrm{NSNS}}
}{0.75\widetilde{M}_{\mathrm{ej}}(a_{\mathrm{BH}}) T_{\mathrm{Gal}}
  f_{\mathrm{tilt}}}.
\end{equation}
Here we also assume that 75\% of the NS-NS merger ejecta form 
$A > 140$ nuclei \citep[but see][for possible
corrections]{2014ApJ...789L..39W}.
The reddish band in Fig.~\ref{fig:cor} shows this relation between
the two rates for our standard model (DD2,
$f_{\mathrm{tilt}}=0.5$) and an average $a_{\mathrm{BH}}$ between 0.5 and
0.7. The two merger rates are anti-correlated: higher 
contributions from NS-NS mergers reduce the ones from NS-BH mergers
and vice versa. Such an anti-correlation also holds for different EoSs but with
distinct shifts because of the
opposite dependences of $\widetilde{M}_{\mathrm{ej,NSNS}}$
and $\widetilde{M}_{\mathrm{ej}}$ on the stiffness of the EoS.
For very soft EoSs (greenish band)
$\widetilde{M}_{\mathrm{ej}}$ may be reduced by
at most a factor five relative to the DD2 case, whereas
$M_{\mathrm{ej,NSNS}}$ should be $\sim$$10^{-2}\,M_{\odot}$.
If NS-BH mergers dominate, their rate limit is correspondingly
higher, while for NS-NS binaries as main source the
relatively large ejecta mass per event leads to tighter rate 
constraints. The opposite trends apply for stiff EoSs
(bluish band), which yield stronger limits for $R_{\mathrm{NSBH}}$
but allow for more NS-NS coalescences. The intermediate (reddish) case
implies that NS-NS mergers as additional r-process source set a strict
upper bound of $\sim$$6\times 10^{-5}\,\mathrm{yr}^{-1}$ for the 
NS-BH merger rate. The bluish and greenish bands represent rather
extreme EoS variations.

The symbols in Fig.~\ref{fig:cor} display ``pessimistic'' (square),
``realistic'' (diamond) and ``optimistic'' (triangle) merger rates
of both event types as compiled in~\citet{2010CQGra..27q3001A}
to estimate the GW detection probability 
by Advanced LIGO and Advanced Virgo. The ``optimistic'' case is
practically ruled out by our estimates because already NS-BH or NS-NS
mergers alone would overproduce r-process nuclei. The ``realistic''
expectations for NS-NS and NS-BH detection rates
(diamond) are marginally consistent with our constraints.

The vertical solid line in Fig.~\ref{fig:cor} marks $R_{\mathrm{NSBH}}$
derived from the observed binary NS population 
\citep{2004ApJ...614L.137K} with thin vertical lines
indicating the corresponding uncertainties. 
Only very stiff EoSs allow for a substantial NS-BH merger rate
that is clearly compatible with the observed double-NS population.
The thick dashed line shows the sum $R_{\mathrm{NSBH}}+
R_{\mathrm{NSNS}}=10^{-4}\,\mathrm{yr}^{-1}$, which roughly
corresponds to a volumetric rate of $10^3\,\mathrm{Gpc^{-3}yr^{-1}}$ 
derived from the observed short-GRB rate and inferred jet opening
angles \citep{2013arXiv1311.2603B}. Here we use a crude conversion
from the volumetric rate to a Galactic merger
rate \citep{2010CQGra..27q3001A} and assume that every NS-BH and
NS-NS merger produces a beamed GRB. Giving up the latter
assumption would result in a higher rate, i.e., a curve shifted
to the upper right. Given the involved uncertainties the rate inferred
from GRBs is roughly compatible with our estimates, which might be 
a hint that the far majority of COMs result in GRBs unless
another, r-process inactive source contributes to making short GRBs.

Any rate estimates, e.g.\ by population synthesis models, should be
checked for consistency with the correlation given by
Eq.~\eqref{eq:cor}. For instance, considering the r-process
yields of NS-NS mergers, the NS-BH merger rates
of~\citet{2012ApJ...759...52D} are roughly compatible with our
constraint. We emphasize that our time-averaged rates are
upper limits of the current merger rate 
\citep[e.g.,][]{2013ApJ...779...72D,2013arXiv1307.0959M,2014arXiv1407.3796S}.
Our rate constraints would even become more stringent if NS-NS and
NS-BH mergers were not the only sources of $A > 140$ r-process
elements \citep[for suggestions, see, 
e.g.,][]{2007PhR...450...97A}.
While eccentric COMs with their higher ejecta masses
\citep[e.g.][]{2013MNRAS.430.2585R} might occur with significant 
frequencies in dense stellar systems \citep{2010ApJ...720..953L},
additional ejecta from massive NSs and BH-torus systems as remnants 
of COMs
\citep[e.g.][]{2013MNRAS.435..502F,2014arXiv1402.4803M,2014arXiv1405.6730P,2014arXiv1406.2687J}
are expected to contribute to the production of $A>140$ r-nuclei only
on a lower level \citep{2014arXiv1406.2687J}.
Conversely, our bounds would be somewhat higher if COMs were less efficient
in enriching the Galactic gas with r-elements, which might be suggested 
by the fact that 50\% of the short GRBs exhibit projected offsets
of $>$1.5 times the half-light radii of their host galaxies
\citep{2013arXiv1311.2603B}.

\section{Summary and outlook}\label{sec:sum}

We quantified an independent upper limit on the NS-BH
merger rate, which, however, depends on the uncertain average BH
spin. Adopting a moderately big spin parameter of $\gtrsim$0.5,
the average merger rate should be $\lesssim$$6\times
10^{-5}\,\mathrm{yr}^{-1}$ with an uncertainty factor of a few.  
We point out
that the NS-BH merger rate is correlated with the NS-NS coalescence
rate, and population synthesis models should be checked for consistency
with our combined constraint. The GW detection rates classified as
``optimistic'' (``realistic'') for Advanced LIGO and Virgo~\citep{2010CQGra..27q3001A} are incompatible (marginally
consistent) with our estimates considering that we deduced time-averaged
rates, which are probably higher than the current rate relevant
for GW detections \citep{2013ApJ...779...72D,2013arXiv1307.0959M};
but see \citet{2003MNRAS.342.1169V} for opposite trends. 
Our rate limits are also roughly consistent with
the possibility that NS-NS and NS-BH mergers are the dominant sources 
of short GRBs.

For improving our estimates, relativistic NS-BH merger simulations
are needed for larger sets of high-density EoSs, different
NS masses and BH spin orientations. Instead of 
time-averaged rate constraints a time-dependent analysis is
desirable, accounting for the Galactic metallicity evolution
and a consistent treatment of the metallicity dependence of NS-NS
and NS-BH populations. More observational data on BH spins and a better 
understanding of natal BH spins and the spin evolution would reduce uncertainties considerably.

\begin{acknowledgments}
We thank T.~Tauris for discussions, M.~Hempel for EoS tables,
and M.~Dominik and C.~Belczynski for population synthesis data and
advice. A.B.\ is Marie Curie
Intra-European Fellow within the 7th European Community Framework
Programme (IEF 331873). S.G is F.R.S.-FNRS Research
Associate. This work was supported by DFG through
grants SFB/TR7 and EXC-153. Computing resources were provided by RZG and LRZ Garching.
\end{acknowledgments}


\end{document}